\documentclass[conference]{IEEEtran}
\IEEEoverridecommandlockouts
\usepackage{amsmath,amssymb,amsfonts}
\usepackage{amsthm}
\usepackage[ruled,linesnumbered]{algorithm2e}
\usepackage{array}
\usepackage{color}
\usepackage{cite}
\usepackage{cuted}
\usepackage{graphicx}
\usepackage{hyperref}
\usepackage{mathrsfs}
\usepackage{multirow}
\usepackage[caption=false,font=scriptsize,labelfont=sf,textfont=sf]{subfig}
\usepackage{threeparttable}
\usepackage{url}
\usepackage{verbatim}

\newcommand\dif{\mathrm{d}}

\newtheorem{theorem}{Theorem}

\def\BibTeX{{\rm B\kern-.05em{\sc i\kern-.025em b}\kern-.08em
    T\kern-.1667em\lower.7ex\hbox{E}\kern-.125emX}}

\begin{document}

\title{Optimal Investment under the Influence of Decision-changing Imitation}

\author{\IEEEauthorblockN{Huisheng Wang and H. Vicky Zhao}
\IEEEauthorblockA{Department of Automation, Tsinghua University, Beijing, China \\
\href{whs22@mails.tsinghua.edu.cn}{whs22@mails.tsinghua.edu.cn}, \href{vzhao@tsinghua.edu.cn}{vzhao@tsinghua.edu.cn}}}

\maketitle
\thispagestyle{plain}

\begin{abstract}
Decision-changing imitation is a prevalent phenomenon in financial markets, where investors imitate others' decision-changing rates when making their own investment decisions. In this work, we study the optimal investment problem under the influence of decision-changing imitation involving one leading expert and one retail investor whose decisions are unilaterally influenced by the leading expert. In the objective functional of the optimal investment problem, we propose the integral disparity to quantify the distance between the two investors' decision-changing rates. Due to the underdetermination of the optimal investment problem, we first derive its general solution using the variational method and find the retail investor's optimal decisions under two special cases of the boundary conditions. We theoretically analyze the asymptotic properties of the optimal decision as the influence of decision-changing imitation approaches infinity, and investigate the impact of decision-changing imitation on the optimal decision. Our analysis is validated using numerical experiments on real stock data. This study is essential to comprehend decision-changing imitation and devise effective mechanisms to guide investors' decisions.
\end{abstract}


\section{Introduction}
With the development of online social media, financial information platforms like Yahoo Finance and Xueqiu have become indispensable tools for investors. These platforms provide retail investors access to real-time market information, as well as opportunities to engage with investment decisions shared by experienced leading experts \cite{bodnaruk2015financial}.  The retail investors' decisions are not only influenced by market information but also by the leading experts \cite{brown2008neighbors}. Therefore, it is crucial to study how leading experts' decisions influence others' decisions in the entire financial market, as this plays a pivotal role in fostering the development of financial markets. 

In financial markets, investors allocate assets that offer high returns and low volatility to maximize their expected utility over the investment period. Markowitz studied static optimal asset allocation at a single time step \cite{1952Portfolio}. Merton extended it to continuous-time dynamic asset allocation over an investment period and formulated the optimal investment problem, known as the Merton problem \cite{merton1969lifetime}. Using stochastic analysis and optimal control, the analytical solutions of investors' optimal decisions under different utility functions were derived in \cite{rogers2013optimal}. 

However, the above works did not consider the influence of leading experts' decisions on those of retail investors. Due to the rich investment experience and extensive influence of leading experts, retail investors' decisions often tend to align with those of leading experts. This phenomenon, known as \textit{herd behavior} in behavioral economics, has been extensively studied in \cite{hu2016study}. Herd behavior suggests that when the leading expert holds \$$N$ in an asset, the retail investors tend to imitate this total holding. Numerous qualitative studies have confirmed the significant impact of herd behavior on retail investors' decisions \cite{yu2019detection, zhou2022internet}, and our prior work in \cite{wang2024herd} quantitatively analyzed the influence of herd behavior among investors based on the optimal investment problem.

In financial markets, a prevalent phenomenon is that the retail investor imitates the changing rate of the leading expert's decision \cite{lakonishok1992impact}, which we call \textit{decision-changing imitation}. Decision-changing imitation suggests that if the leading expert increases or decreases his/her investment in a risky asset by \$$\Delta N$ during a time interval $\Delta t$, the retail investor will likewise adjust their investment by \$$\Delta N$ in the same time interval. We call \$${\Delta N}/{\Delta t}$ the investor's decision-changing rate. Compared to herd behavior, the investigation of decision-changing imitation is supported by two rationales. First, due to confidentiality concerns, leading experts are reluctant to publicly share their total holdings on financial information platforms, making it challenging for retail investors to imitate their decisions directly. Conversely, retail investors often have access to information regarding the changing rate of the leading experts' decisions \cite{stotz2012retail}. Second, most leading experts are wealthy investors with substantial holdings, making it financially burdensome for retail investors to imitate their decisions directly \cite{belak2022optimal}. On the contrary, imitating the decision-changing rate only involves adjusting holdings, which is relatively more feasible for retail investors.

To the best of our knowledge, few works have quantitatively studied the impact of decision-changing imitation on investors' optimal investment decisions. In this work, we address this gap by formulating an optimal investment problem under the influence of decision-changing imitation, which involves one leading expert and one retail investor whose decisions are influenced by the former. We then quantitatively analyze how decision-changing imitation affects the decision-making process of the retail investor.


The structure of this paper is as follows. We formulate the optimal investment problem in Section \ref{sec:decision}. We derive the general solution to the optimal investment problem and derive the retail investor's optimal decisions under two special cases of the boundary conditions in Section \ref{sec:solution}. In Section \ref{sec:analysis}, we theoretically analyze the asymptotic properties of the retail investor's optimal decision as the influence intensity of decision-changing imitation approaches infinity, and investigate the impact of decision-changing imitation on the optimal decision. In Section \ref{sec:simulation}, we conduct numerical experiments on real stock data to validate our analysis. Section \ref{sec:conclusion} is the conclusion.

\section{Problem Formulation}
\label{sec:decision}
We consider the scenario involving one retail investor $A_1$ and one leading expert $A_2$, and the retail investor's decisions are unilaterally influenced by the leading expert.  Following the prior work in \cite{merton1969lifetime}, we assume that the two investors $A_i$ ($i=1,2$) invest in the period $\mathcal{T}:=[0, T]$ in a financial market consisting of a risk-free asset and a risky asset. We use the compound interest pricing model to represent the price process of the risk-free asset with fixed income, and use the geometric Brownian motion to represent the price process of the risky asset with volatile returns, which are two common pricing models in optimal investment \cite{merton1969lifetime}. Let $r$ denote the interest rate of the risk-free asset, and let $v$ and $\sigma$ denote the excess return rate and the volatility of the risky asset, respectively. Let $x_i$ denote the initial wealth of $A_i$, and let $\{P_i(t)\}_{t\in\mathcal{T}}$ denote the holding of the risky asset held by $A_i$. From \cite{merton1969lifetime}, the wealth process $\{X_i(t)\}_{t\in\mathcal{T}}$ of $A_i$ can be expressed as 
\begin{equation}
    \dif X_i(t)=[rX_i(t)+vP_i(t)]\dif t+\sigma P_i(t)\dif W(t),
    \label{eq:sde}
\end{equation}
subject to $X_i(0)=x_i$, where $\{W(t)\}_{t\in\mathcal{T}}$ represents a standard Brownian motion.

Given the above market setting, we then formulate the optimal investment problems for the leading expert and the retail investor, respectively.

\subsection{Optimal Investment Problem for the Leading Expert}
Note that the retail investor does not influence the leading expert's decisions. Following the prior work in \cite{merton1969lifetime}, the leading expert determines $\{P_2(t)\}_{t\in\mathcal{T}}$ to maximize his/her expected utility of the terminal wealth $\mathbb{E}\phi_2(X_2(T))$. The utility function $\phi_i(X_i(T))$ satisfies the characteristics of diminishing marginal returns and concavity. In this work, we consider the  Constant Absolute Risk Aversion utility \cite{pratt1978risk}, which is
\begin{equation}
    \phi_i(X_i(T))=-\frac{1}{\alpha_i}\mathrm{e}^{-\alpha_iX_i(T)},
    \label{eq:utility}
\end{equation}
where $\alpha_i>0$ is called the \textit{risk aversion coefficient} of $A_i$. A larger $\alpha_i$ means that $A_i$ is less risk-seeking, and the utility becomes more sensitive to changes in the terminal wealth. 

Therefore, the optimal investment problem for $A_2$ becomes
\begin{align*}
    \text{\textbf{Problem 1.}}&\sup_{P_2\in\mathcal{U}}\mathbb{E}\phi_2(X_2(T))\\
    \text{s.t.\quad}&\dif X_2(t)=[rX_2(t)+vP_2(t)]\dif t+\sigma P_2(t)\dif W(t),\\
    &X_2(0)=x_2,
\end{align*}
which is the Merton problem \cite{merton1969lifetime}. In Problem 1, $\mathcal{U}$ represents the set of admissible decisions, which is a subset of $\mathcal{L}^1(\mathcal{T}):=\left\{u \middle|\mathbb{E}\int_0^T|u(t)|\dif t<\infty\right\}$. We further assume that $\mathcal{U}\subset\mathcal{C}^1(\mathcal{T})$, i.e., the decision is continuously differentiable. 

From \cite{merton1969lifetime}, the leading expert's optimal decision is
\begin{equation}
    \bar{P}_2(t)=\frac{v}{\alpha_2\sigma^2}\mathrm{e}^{r(t-T)},t\in\mathcal{T}.
    \label{eq:merton-optimal-decision}
\end{equation}
We call \eqref{eq:merton-optimal-decision} the \textit{rational decision} without considering other's influence. From \eqref{eq:merton-optimal-decision}, the rational decision is proportional to the excess return rate $v$ and inversely proportional to the volatility $\sigma$ and the risk aversion coefficient $\alpha_2$.

\subsection{Optimal Investment Problem for the Retail Investor}
Next, we formulate the optimal investment problem for the retail investor with the decision-changing imitation. We assume that $A_2$'s decision follows the form of the rational decision $\{\bar{P}_2(t)\}_{t\in\mathcal{T}}$ as in \eqref{eq:merton-optimal-decision}, and $A_1$ has access to $\{\bar{P}_2(t)\}_{t\in\mathcal{T}}$, which is his/her prior knowledge before the investment period. 

Considering the decision-changing imitation, $A_1$ aims to maximize his/her expected utility of the terminal wealth while minimizing the distance between their decision-changing rates. Given a time interval $\Delta t$, $A_1$'s decision-changing rate from time $t$ to $t+\Delta t$ is $\frac{P_1(t+\Delta t)-P_1(t)}{\Delta t}$. Because $\mathcal{U}\subseteq\mathcal{C}^1(\mathcal{T})$, the derivative of $\{P_1(t)\}_{t\in\mathcal{T}}$ always exists. If $\Delta t$ is sufficiently small, we can further replace $\frac{P_1(t+\Delta t)-P_1(t)}{\Delta t}$ with the derivative of the decision, which is
\begin{equation}
    \lim_{\Delta t\to0^+}\frac{P_1(t+\Delta t)-P_1(t)}{\Delta t}=\frac{\dif P(t)}{\dif t}:=\dot{P}_1(t).
\end{equation}
The prior work in \cite{merigo2011induced} used the Euclidean distance to measure the distance between two decisions, i.e., $\frac{1}{2}\int_0^T[P_1(t)-\bar{P}_2(t)]^2\dif t$. Inspired by this, in this work, we define the Euclidean distance between the derivatives of the two investors' decisions $\{\dot{P}_1(t)\}_{t\in\mathcal{T}}$ and $\{\dot{\bar{P}}_2(t)\}_{t\in\mathcal{T}}$ as the \textit{integral disparity}, which is
\begin{equation}
    D(\dot{P}_1,\dot{\bar{P}}_2):=\frac{1}{2}\int_0^T[\dot{P}_1(t)-\dot{\bar{P}}_2(t)]^2\dif t\geqslant0.
    \label{eq:herd-cost}
\end{equation}


Therefore, the optimal investment problem for $A_1$ becomes
\begin{align*}
    \text{\textbf{Problem 2.}}&\sup_{P_1\in\mathcal{U}}\mathbb{E}\phi_1(X_1(T))-\theta D(\dot{P}_1,\dot{\bar{P}}_2)\\
    \text{s.t.\quad}&\dif X_1(t)=[rX_1(t)+vP_1(t)]\dif t+\sigma P_1(t)\dif W(t),\\
    &X_1(0)=x_1,
\end{align*}
where the imitation coefficient $\theta>0$ is to address the tradeoff between the two different objectives, i.e., maximizing the expected utility of the terminal wealth $\mathbb{E}\phi_1(X_1(T))$ and minimizing the integral disparity $D(\dot{P}_1,\dot{\bar{P}}_2)$. When $\theta=0$, the retail investor's optimal decision is his/her rational decision
\begin{equation}
    \bar{P}_1(t)=\frac{v}{\alpha_1\sigma^2}\mathrm{e}^{r(t-T)},t\in\mathcal{T}.
    \label{eq:barP1}
\end{equation}

Note that in Problem 2, the wealth process $\{X_1(t)\}_{t\in\mathcal{T}}$ follows a first-order stochastic differential equation, and we stipulate an initial condition $X_1(0)=x_1$. Similarly, within the objective functional, there exists a first-order derivative $\{\dot{P}_1(t)\}_{t\in\mathcal{T}}$. According to the variational method, two boundary conditions are required to ensure that Problem 2 is well-defined. The specific forms of the two boundary conditions are dependent on the initial and terminal decisions and will be discussed in the next section.


\section{Solution to the Optimal Investment Problem}
\label{sec:solution}
In this section, we first find the retail investor's general optimal decision without boundary conditions. Then, we establish the boundary conditions using the variational method and determine the optimal decision under two special cases of the boundary conditions. Finally, we provide a numerical method to determine the parameters in the optimal decision. All proofs are in the supplementary file \cite{supp}.


\subsection{General solution to the Optimal Decision}
\begin{theorem}\label{th:1}
    The general solution of Problem 2 is
    \begin{align}
        P_1^*(t)=\ &\gamma_1\mathrm{I}_0(\zeta\mathrm{e}^{-rt})+\gamma_2\mathrm{K}_0(\zeta\mathrm{e}^{-rt})\notag\\
        &+\mathrm{I}_0(\zeta\mathrm{e}^{-rt})\mathfrak{K}(\zeta\mathrm{e}^{-rt})-\mathrm{K}_0(\zeta\mathrm{e}^{-rt})\mathfrak{I}(\zeta\mathrm{e}^{-rt}),
        \label{eq:optimal-decision}
    \end{align}
    for $t\in\mathcal{T}$, where 
    \begin{align}
        \zeta:=\ &\frac{\sigma\mathrm{e}^{rT}}{r}\sqrt{\frac{\eta\alpha_1}{\theta}}\quad\text{and}\label{eq:para}\\
        \eta:=\ &\exp\left\{-\alpha_1x\mathrm{e}^{rT}-\alpha_1v\int_0^T\mathrm{e}^{r(T-t)}P_1^*(t)\dif t\right.\notag\\
        &\left.+\ \frac{\alpha_1^2\sigma^2}{2}\int_0^T\mathrm{e}^{2r(T-t)}P_1^{*2}(t)\dif t\right\}\label{eq:int}
    \end{align}
    are two integral constants, $\mathrm{I}_0(\cdot)$ and $\mathrm{K}_0(\cdot)$ represent the zeroth-order modified Bessel and Neumann functions,
    \begin{align}
        \mathfrak{I}(x):=\ &\int_1^x\mathrm{I}_0(y)\left(\frac{\zeta v\mathrm{e}^{-rT}}{\alpha_2\sigma^2y^2}-\frac{\eta v\mathrm{e}^{rT}}{\zeta r^2\theta}\right)\dif y,\quad \text{and}\label{eq:special-i}\\
        \mathfrak{K}(x):=\ &\int_1^x\mathrm{K}_0(y)\left(\frac{\zeta v\mathrm{e}^{-rT}}{\alpha_2\sigma^2y^2}-\frac{\eta v\mathrm{e}^{rT}}{\zeta r^2\theta}\right)\dif y.\label{eq:special-k}
    \end{align}
    In \eqref{eq:optimal-decision}, $\gamma_1,\gamma_2\in\mathbf{R}$ are two parameters satisfying
    \begin{equation}
        [\dot{P}_1^*(0)-\dot{\bar{P}}_2(0)]\delta P_1^*(0)=[\dot{P}_1^*(T)-\dot{\bar{P}}_2(T)]\delta P_1^*(T),
        \label{eq:boundary-cond}
    \end{equation}
    where $\{\delta P_1^*(t)\}_{t\in\mathcal{T}}$ refers to the variation with respect to $\{P_1^*(t)\}_{t\in\mathcal{T}}$, which is a function in $\mathcal{U}$ and represents a small change in the function $\{P_1^*(t)\}_{t\in\mathcal{T}}$. If $P_1^*(t)$ takes a fixed value at time $t$, then $\delta P_1^*(t)=0$ because there is no change at that time, otherwise $\delta P_1^*(t)$ can take any value. $\dot{\bar{P}}_2(0)$ and $\dot{\bar{P}}_2(T)$ are the derivatives of the leading expert's ration decision when $t=0$ and $t=T$, which can be calculated using \eqref{eq:merton-optimal-decision}.
\end{theorem}

Theorem \ref{th:1} provides the general solution to Problem 2. However, to obtain the retail investor's optimal decision, there remain two problems. First, we need to establish the forms of the boundary conditions using \eqref{eq:boundary-cond} and determine the parameters $\gamma_1$ and $\gamma_2$ according to the boundary conditions. Second, because \eqref{eq:para} and \eqref{eq:int} do not have closed-form solutions, we need to numerically calculate the integral constants $\zeta$ and $\eta$. In the following, we address these problems respectively.

\subsection{Two Special Cases of the Boundary Conditions}
Note that when \eqref{eq:boundary-cond} holds, due to the arbitrariness of $\delta P_1^*(0)$ and $\delta P_1^*(T)$, both the left and right-hand sides of \eqref{eq:boundary-cond} must be equal to zero. 
That is, when $t=0$ and $t=T$, if $P_1^*(t)$ is not a fixed value, then $\dot{P}_1^*(t)$ must equal $\dot{\bar{P}}_2(t)$. Therefore, the boundary conditions include four possible cases:
\begin{itemize}
    \item Case 1: $\delta P_1^*(0)=0$ and $\delta P_1^*(T)=0$,
    \item Case 2: $\dot{P}_1^*(0)=\dot{\bar{P}}_2(0)$ and $\dot{P}_1^*(T)=\dot{\bar{P}}_2(T)$,
    \item Case 3: $\delta P_1^*(0)=0$ and $\dot{P}_1^*(T)=\dot{\bar{P}}_2(T)$, and
    \item Case 4: $\dot{P}_1^*(0)=\dot{\bar{P}}_2(0)$ and $\delta P_1^*(T)=0$,
\end{itemize}
where $\delta P_1^*(0)=0$ and $\delta P_1^*(T)=0$ mean that $P_1^*(0)$ and $P_1^*(T)$ takes two fixed values that need to be determined. 

The four cases of boundary conditions correspond to four different types of retail investors in real financial markets, depending on whether they use fixed values for their decisions at $t=0$ and $t=T$. In Case 1, they fix $P_1^*(0)$ and $P_1^*(T)$. In Case 2, they do not fix their decisions at these times, and thus, from the above analysis, they have to choose $\dot{P}_1^*(0)=\dot{\bar{P}}_2(0)$ and $\dot{P}_1^*(T)=\dot{\bar{P}}_2(T)$ for \eqref{eq:boundary-cond} to hold. Case 3 and Case 4 correspond to the scenarios where they only fix $P_1^*(0)$ and $P_1^*(T)$, respectively. 

In the following, we focus on Cases 1 and 2 to gain valuable insights. These two cases offer clear physical interpretations, and their solutions are relatively straightforward. The solutions for Case 3 and Case 4 are similar and omitted here.

\subsubsection{Case 1}
In this case, we need to determine the values of $P_1^*(0)$ and $P_1^*(T)$. The prior works in \cite{brown1969prior, tversky1991loss} suggest that investors tend to make more rational decisions at the initial and terminal times than at other times. Inspired by this, here, we assume that when $t=0$ and $t=T$, the retail investor's decision is equal to his/her rational decision. Note that if we set other values for $P_1^*(0)$ and $P_1^*(T)$, we can use a similar method to solve. From \eqref{eq:barP1}, the two boundary conditions are
\begin{equation}
    P_1^*(0)=\frac{v}{\alpha_1\sigma^2}\mathrm{e}^{-rT}\quad\text{and}\quad
    P_1^*(T)=\frac{v}{\alpha_1\sigma^2}.
    \label{eq:termdir}
\end{equation}

When \eqref{eq:termdir} hold, Problem 2 becomes
\begin{align*}
    \text{\textbf{Problem 3.}}&\sup_{P_1\in\mathcal{U}}\mathbb{E}\phi_1(X_1(T))-\theta D(\dot{P}_1,\dot{\bar{P}}_2)\\
    \text{s.t.\quad}&\dif X_1(t)=[rX_1(t)+vP_1(t)]\dif t+\sigma P_1(t)\dif W(t),\\
    &X_1(0)=x_1,\\
    &P_1(0)=\frac{v}{\alpha_1\sigma^2}\mathrm{e}^{-rT},\quad P_1(T)=\frac{v}{\alpha_1\sigma^2}.
\end{align*}

Problem 3 describes the scenario where the retail investor uses his/her rational decisions at the beginning and the end of the investment period but is subject to the influence of decision-changing imitation at all other times.

Next, we determine the parameters $\gamma_1$ and $\gamma_2$ of $\{P_1^*(t)\}_{t\in\mathcal{T}}$ in Problem 3. To simplify the notation, we denote
\begin{equation}
    \begin{cases}
    \begin{aligned}
        \iota_0^0&:=\mathrm{I}_0(\zeta),&\iota_0^1&:=\mathrm{I}_0(\xi),&\iota_1^0&:=\mathrm{I}_1(\zeta),&\iota_1^1&:=\mathrm{I}_1(\xi),\\
        \kappa_0^0&:=\mathrm{K}_0(\zeta),&\kappa_0^1&:=\mathrm{K}_0(\xi),&\kappa_1^0&:=\mathrm{K}_1(\zeta),&\kappa_1^1&:=\mathrm{K}_1(\xi),\\
        \tilde\iota^0&:=\mathfrak{I}(\zeta),&\tilde\iota^1&:=\mathfrak{I}(\xi),&\tilde\kappa^0&:=\mathfrak{K}(\zeta),&\tilde\kappa^1&:=\mathfrak{K}(\xi),\\
        \tilde\iota&:=\tilde\iota^1-\tilde\iota^0,&\tilde\kappa&:=\tilde\kappa^1-\tilde\kappa^0,
    \end{aligned}
    \end{cases}
    \label{eq:hyp}
\end{equation}
where $\xi:=\zeta\mathrm{e}^{-rT}$, and $\mathrm{I}_1(\cdot)$ and $\mathrm{K}_1(\cdot)$ represent the first-order modified Bessel and Neumann functions, respectively.

\begin{theorem}\label{th:2}
    The parameters $\gamma_1$ and $\gamma_2$ of $A_1$'s optimal decision $\{P_1^*(t)\}_{t\in\mathcal{T}}$ in Problem 3 are
    \begin{align}
        \gamma_1=\ &-\tilde{\kappa}^0-\frac{\kappa_0^0(\kappa_0^1\tilde{\iota}-\iota_0^1\tilde{\kappa})+(\kappa_0^0-\kappa_0^1\mathrm{e}^{-rT})\frac{v}{\alpha_1\sigma^2}}{\iota_0^0\kappa_0^1-\iota_0^1\kappa_0^0},\notag\\
        \gamma_2=\ &\tilde{\iota}^0+\frac{\iota_0^0(\kappa_0^1\tilde{\iota}-\iota_0^1\tilde{\kappa})+(\iota_0^0-\iota_0^1\mathrm{e}^{-rT})\frac{v}{\alpha_1\sigma^2}}{\iota_0^0\kappa_0^1-\iota_0^1\kappa_0^0}.
        \label{eq:case12}
    \end{align}
\end{theorem}

\begin{algorithm}[t]
\footnotesize
	\caption{Iterative Method for $\zeta$ and $\eta$.}
	\label{alg:1}
	\KwIn{Interest rate: $r$; Excess return rate: $v$; Volatility: $\sigma$; Initial wealth: $x_1$; Risk aversion coefficients: $\alpha_1,\alpha_2$; Investment period: $T$; Imitation coefficient: $\theta$; Initial integral constants: $\zeta_0,\eta_0$; Tolerance: $\varepsilon$.}
	\KwOut{Integral constants: $\zeta$ and $\eta$.}  
	$\zeta^{(0)}=\zeta_0$, $\eta^{(0)}=\eta_0$, $\Delta \zeta^{(0)}=\Delta \eta^{(0)}=+\infty$, $k=0$;

	\While{$\Delta\zeta^{(k)}\geqslant\varepsilon$ \textnormal{\textbf{or}} $\Delta\eta^{(k)}\geqslant\varepsilon$}{
    \If{Case 1}
    {Calculate ${\iota_0^{0}}^{(k)}$, ${\iota_1^{0}}^{(k)}$, ${\kappa_0^{0}}^{(k)}$, ${\kappa_1^{0}}^{(k)}$, $\tilde{\iota}^{0(k)}$, $\tilde{\kappa}^{0(k)}$, $\tilde{\iota}^{(k)}$, and $\tilde{\kappa}^{(k)}$ using \eqref{eq:hyp};
    
    Calculate $\gamma_1^{(k)}$ and $\gamma_2^{(k)}$ using \eqref{eq:case12};}
    
    \If{Case 2}
    {Calculate ${\iota_0^{1}}^{(k)}$, ${\iota_1^{1}}^{(k)}$, ${\kappa_0^{1}}^{(k)}$, ${\kappa_1^{1}}^{(k)}$, $\tilde{\iota}^{1(k)}$, $\tilde{\kappa}^{1(k)}$, $\tilde{\iota}^{(k)}$, and $\tilde{\kappa}^{(k)}$ using \eqref{eq:hyp};
    
    Calculate $\gamma_1^{(k)}$ and $\gamma_2^{(k)}$ using \eqref{eq:case22};}
    
    Calculate $P_1^{*(k)}$ using \eqref{eq:optimal-decision};
    
    Calculate $\eta^{(k+1)}$ using \eqref{eq:int};
    
    Calculate $\zeta^{(k+1)}$ using \eqref{eq:para};
    
    $\Delta\zeta^{(k+1)}=|\zeta^{(k+1)}-\zeta^{(k)}|$, $\Delta\eta^{(k+1)}=|\eta^{(k+1)}-\eta^{(k)}|$;
    
    $k\leftarrow k+1$;}
    
    $\zeta\approx\zeta^{(k)}$ ($|\zeta-\zeta^{(k)}|<\varepsilon$), $\eta\approx\eta^{(k)}$ ($|\eta-\eta^{(k)}|<\varepsilon$).
\end{algorithm}

\subsubsection{Case 2}
In this case, we do not need to set the values of $P_1^*(0)$ and $P_1^*(T)$ as in Case 1, but let $\dot{P}_1^*(0)=\dot{\bar{P}}_2(0)$ and $\dot{P}_1^*(T)=\dot{\bar{P}}_2(T)$. Then Problem 2 becomes
\begin{align*}
    \text{\textbf{Problem 4.}}&\sup_{P_1\in\mathcal{U}}\mathbb{E}\phi_1(X_1(T))-\theta D(\dot{P}_1,\dot{\bar{P}}_2)\\
    \text{s.t.\quad}&\dif X_1(t)=[rX_1(t)+vP_1(t)]\dif t+\sigma P_1(t)\dif W(t),\\
    &X_1(0)=x_1,\\
    &\dot{P}_1(0)=\frac{rv}{\alpha_2\sigma^2}\mathrm{e}^{-rT},\quad \dot{P}_1(T)=\frac{rv}{\alpha_2\sigma^2}.
\end{align*}

Problem 4 describes the scenario where the retail investor does not fix his/her initial or terminal decision and is subject to the influence of decision-changing imitation over the investment period.
The parameters $\gamma_1$ and $\gamma_2$ of $\{P_1^*(t)\}_{t\in\mathcal{T}}$ in Problem 4 can be calculated using Theorem \ref{th:2+}.
\begin{theorem}\label{th:2+}
    The parameters $\gamma_1$ and $\gamma_2$ of $A_1$'s optimal decision $\{P_1^*(t)\}_{t\in\mathcal{T}}$ in Problem 4 are
    \begin{align}
        \gamma_1=\ &-\tilde{\kappa}^0+\frac{\kappa_1^0(\kappa_1^1\tilde{\iota}+\iota_1^1\tilde{\kappa})+(\kappa_1^0\mathrm{e}^{rT}-\kappa_1^1\mathrm{e}^{-rT})\frac{v}{\zeta\alpha_2\sigma^2}}{\iota_1^0\kappa_1^1-\iota_1^1\kappa_1^0},\notag\\
        \gamma_2=\ &\tilde{\iota}^0+\frac{\iota_1^0(\kappa_1^1\tilde{\iota}+\iota_1^1\tilde{\kappa})+(\iota_1^0\mathrm{e}^{rT}-\iota_1^1\mathrm{e}^{-rT})\frac{v}{\zeta\alpha_2\sigma^2}}{\iota_1^0\kappa_1^1-\iota_1^1\kappa_1^0}.
        \label{eq:case22}
    \end{align}
\end{theorem}

Note that when $\theta=0$, the optimal decision in Problem 4 is incompatible with the boundary conditions. We address this problem in the supplementary file.

\subsection{Integral Constants}
Next, we calculate the integral constants $\zeta$ and $\eta$ in Theorem \ref{th:1} using the iterative method in Algorithm \ref{alg:1}. The initial values $\zeta_0$ and $\eta_0$ of the iterative method affect the convergence and need to be selected appropriately according to different parameters. In this work, given the parameters in Section \ref{sec:simulation}, we set $\zeta_0=\eta_0=1$ and find that Algorithm \ref{alg:1} is convergent. For other parameters, we can use the numerical solver in Python to find $\zeta$ and $\eta$. Details are in the supplementary file.

\subsection{Summary}
Given the parameters $\gamma_1$ and $\gamma_2$ and the integral constants $\zeta$ and $\eta$, the retail investor's optimal decision $\{P_1^*(t)\}_{t\in\mathcal{T}}$ can be calculated using Theorem \ref{th:1}--\ref{th:2+}. 

\section{Asymptotic Properties of the Optimal Decision}
\label{sec:analysis}
In this section, we study the influence of decision-changing imitation on the retail investor's optimal decision. As mentioned above, we use the imitation coefficient $\theta$ to quantify the intensity of decision-changing imitation, and a larger $\theta$ means that the influence of the leading expert on the retail investor's decision is stronger. Note that the optimal decision in \eqref{eq:optimal-decision} is complex to analyze. To gain insights, here, we focus on the special scenario when $\theta$ approaches infinity, and in Section \ref{sec:simulation}, we run simulations for the general cases when $\theta\in(0,+\infty)$. We denote the retail investor's optimal decision when $\theta$ approaches infinity as $\{P_{1\infty}^*(t)\}_{t\in\mathcal{T}}$, which we call the \textit{asymptotic decision}. 
 
\subsection{Case 1}
In Problem 3, the retail investor's asymptotic decision can be calculated using Theorem \ref{th:3}.
\begin{theorem}\label{th:3}
    $A_1$'s asymptotic decision $\{P_{1\infty}^*(t)\}_{t\in\mathcal{T}}$ in Problem 3 is
    \begin{align}
        P_{1\infty}^*(t)=\ &\bar{P}_2(t)+\frac{v(\alpha_1^{-1}-\alpha_2^{-1})}{\sigma^2}\cdot\frac{1-\mathrm{e}^{-rT}}{T}\cdot t\notag\\
        &+\frac{v(\alpha_1^{-1}-\alpha_2^{-1})}{\sigma^2}\cdot\mathrm{e}^{-rT},\forall t\in\mathcal{T}.
        \label{eq:sol-case1}
    \end{align}
\end{theorem}

To theoretically analyze the influence of the leading expert's decision on that of the retail investor, due to the complexity of \eqref{eq:sol-case1}, we cannot use the rational decision decomposition method in \cite{wang2024herd} to quantify the relationship in magnitude between the retail investor's asymptotic decision and the two agents' rational decisions. Therefore, in the following, we directly compare the retail investor's asymptotic decision $\{P_{1\infty}^*(t)\}_{t\in\mathcal{T}}$ with the two agents' rational decisions $\{\bar{P}_1(t)\}_{t\in\mathcal{T}}$ and $\{\bar{P}_2(t)\}_{t\in\mathcal{T}}$, as shown in Theorem \ref{co:1}.
\begin{theorem}\label{co:1}
In Problem 3, when $t\in\mathcal{T}$, we have
\begin{align}
    &P_{1\infty}^*(t)\geqslant\bar{P}_1(t)>\bar{P}_2(t) \quad \text{if} \quad \alpha_1<\alpha_2,\quad\text{and}\label{eq:co-11}\\
    &P_{1\infty}^*(t)\leqslant\bar{P}_1(t)<\bar{P}_2(t) \quad \text{if} \quad \alpha_1>\alpha_2.
    \label{eq:co-12}
\end{align}
\end{theorem}


Theorem \ref{co:1} demonstrates that in Case 1, under the influence of decision-changing imitation, the retail investor's asymptotic decision $\{P_{1\infty}^*(t)\}_{t\in\mathcal{T}}$ diverges more from the leading expert's rational decision $\{\bar{P}_2(t)\}_{t\in\mathcal{T}}$ compared to his/her rational decision $\{\bar{P}_1(t)\}_{t\in\mathcal{T}}$, i.e., $|P_{1\infty}^*(t)-\bar{P}_2(t)|>|P_{1\infty}^*(t)-\bar{P}_1(t)|$ for $t\in\mathcal{T}$. 

When $\alpha_1<\alpha_2$, the retail investor prefers higher risk than the leading expert. If the retail investor sets his/her initial and terminal decisions as the rational decisions, from Theorem \ref{co:1}, we have $P_{1\infty}^*(t)\geqslant\bar{P}_1(t)$, which shows that he/she becomes more risk-seeking with consideration of the decision-changing imitation when compared to his/her rational decision. This is because, given the boundary conditions \eqref{eq:termdir}, it can be easily proved that $D(\dot{P}_{1\infty}^*,\dot{\bar{P}}_2)\ne0$. As $\theta$ approaches infinity and given that $\mathbb{E}\phi_1(X_1(T))<\infty$, Problem 3 is equivalent to minimize $D(\dot{P}_{1\infty}^*,\dot{\bar{P}}_2)$. Therefore, we must have $\ddot{P}_{1\infty}^*(t)=\ddot{\bar{P}}_2(t)$, which is proved in the supplementary file. That is, to minimize the distance of the decision-changing rates, we must ensure their accelerations are the same. When $\alpha_1<\alpha_2$, from \eqref{eq:merton-optimal-decision} and \eqref{eq:barP1}, we have $\ddot{\bar{P}}_1(t)>\ddot{\bar{P}}_2(t)=\ddot{P}_{1\infty}^*(t)>0$, i.e., the decisions are convex functions. Given $P_{1\infty}^*(0)=\bar{P}_1(0)$ and $P_{1\infty}^*(T)=\bar{P}_1(T)$, we have $P_{1\infty}^*(t)\geqslant\bar{P}_1(t)$ and thus \eqref{eq:co-11}. 

For the case when $\alpha_1>\alpha_2$, i.e., the retail investor prefers lower risk than the leading expert, we can draw a similar conclusion, and show that the retail investor becomes less risk-seeking with consideration of the decision-changing imitation when compared to his/her rational decision.


\subsection{Case 2}
In Problem 4, the retail investor's asymptotic decision can be calculated using Theorem \ref{th:4}.
\begin{theorem}\label{th:4}
    $A_1$'s asymptotic decision $\{P_{1\infty}^*(t)\}_{t\in\mathcal{T}}$ in Problem 4 is
    \begin{equation}
        P_{1\infty}^*(t)=\bar{P}_2(t)+\frac{2v(\alpha_1^{-1}-\alpha_2^{-1})}{\sigma^2}\cdot\frac{\mathrm{e}^{rT}-1}{\mathrm{e}^{2rT}-1},\forall t\in[0,T].
    \label{eq:sol-case2}
    \end{equation}
\end{theorem}

Furthermore, we have the following Theorem \ref{co:2}.
\begin{theorem}\label{co:2}
In Problem 4, when $t\in[0,\tau]$, we have
\begin{align}
    &P_{1\infty}^*(t)\geqslant\bar{P}_1(t)>\bar{P}_2(t) \quad \text{if} \quad \alpha_1<\alpha_2,\quad\text{and}\label{eq:co-31}\\
    &P_{1\infty}^*(t)\leqslant\bar{P}_1(t)<\bar{P}_2(t) \quad \text{if} \quad \alpha_1>\alpha_2,\label{eq:co-32}
\end{align}
and when $t\in[\tau,T]$, we have
\begin{align}
    &\bar{P}_1(t)\geqslant P_{1\infty}^*(t)>\bar{P}_2(t) \quad \text{if} \quad \alpha_1<\alpha_2,\quad\text{and}\label{eq:co-33}\\
    &\bar{P}_1(t)\leqslant P_{1\infty}^*(t)<\bar{P}_2(t) \quad \text{if} \quad \alpha_1>\alpha_2,\label{eq:co-34}
\end{align}
where
\begin{equation}
    \tau=T+\frac{1}{r}\ln\frac{\mathrm{e}^{rT}-1}{\mathrm{e}^{2rT}-1}+\frac{\ln2}{r}\in[0,T].
    \label{eq:tau}
\end{equation}
\end{theorem}


Theorem \ref{co:2} demonstrates that in Case 2, under the influence of decision-changing imitation, when $t<\tau$, the retail investor's asymptotic decision $\{P_{1\infty}^*(t)\}_{t\in\mathcal{T}}$ diverges more from the leading expert's rational decision $\{\bar{P}_2(t)\}_{t\in\mathcal{T}}$ compared to his/her rational decision $\{\bar{P}_1(t)\}_{t\in\mathcal{T}}$, i.e., $|P_{1\infty}^*(t)-\bar{P}_2(t)|>|P_{1\infty}^*(t)-\bar{P}_1(t)|$. However, when $t>\tau$, $\{P_{1\infty}^*(t)\}_{t\in\mathcal{T}}$ falls between $\{\bar{P}_1(t)\}_{t\in\mathcal{T}}$ and $\{\bar{P}_2(t)\}_{t\in\mathcal{T}}$. 

In the supplementary file, we prove that given $\dot{P}_1^*(0)=\dot{\bar{P}}_2(0)$ and $\dot{P}_1^*(T)=\dot{\bar{P}}_2(T)$, the integral disparity reaches its infimum, i.e., $D(\dot{P}_{1\infty}^*,\bar{P}_2)=0$, when $\dot{P}_{1\infty}^*(t)=\dot{\bar{P}}_2(t)$ for $t\in\mathcal{T}=[0,T]$. Therefore, $\{P_{1\infty}^*(t)\}_{t\in\mathcal{T}}$ equals $\{\bar{P}_2(t)\}_{t\in\mathcal{T}}$ plus an offset. As $\theta$ approaches infinity and given that $\mathbb{E}\phi_1(X_1(T))<\infty$, Problem 4 is equivalent to maximize $\mathbb{E}\phi_1(X_1(T)$. If $P_{1\infty}^*(t)\ne\bar{P}_1(t)$ for all $t\in\mathcal{T}$, it can be easily proved that the expected utility $\mathbb{E}\phi_1(X_1(T))$ cannot reach its supremum. Therefore, there must exist $\tau\in\mathcal{T}$ such that $P_{1\infty}^*(\tau)=\bar{P}_1(\tau)$, from which we can determine the value of $\tau$ as in \eqref{eq:tau}. When $\alpha_1<\alpha_2$, given $\dot{P}_{1\infty}^*(t)=\dot{\bar{P}}_2(t)<\dot{\bar{P}}_1(t)$ for $t\in\mathcal{T}$ and $P_{1\infty}^*(\tau)=\bar{P}_1(\tau)$, according to the property of the continuously differentiable function, we must have $P_{1\infty}^*(t)\geqslant\bar{P}_1(t)$ when $t\in[0,\tau]$, and $P_{1\infty}^*(t)\leqslant\bar{P}_1(t)$ when $t\in[\tau,T]$. Therefore, we have \eqref{eq:co-31} and \eqref{eq:co-32}. 
That is, the retail investor prefers higher risk than the leading expert, the retail investor becomes more risk-seeking during the earlier part of the investment period and less risk-seeking during the latter part with the decision-changing imitation.

For the case when $\alpha_1<\alpha_2$, i.e., the retail investor prefers lower risk than the leading expert, we can draw a similar conclusion, and show that the retail investor becomes less risk-seeking during the earlier part of the investment period and more risk-seeking during the latter part with the decision-changing imitation.



\section{Numerical Experiments}
\label{sec:simulation}
In this section, we conduct numerical experiments on real stock data to validate our analysis in Section \ref{sec:solution} and \ref{sec:analysis}, and analyze the influence of decision-changing imitation on the retail investor’s optimal decision when $\theta\in(0,+\infty)$.

\subsection{Parameter Settings}

Over fifty years, from January 1974 to December 2023, denoted as $T=50$, we gather daily closing prices of the Dow Jones Industrial Average, serving as a proxy for the risky asset's prices, and estimate its excess return rate $v$ to be $0.03$ and volatility $\sigma$ to be $0.17$. We obtain the risk-free interest rate $r$, approximated at $0.04$, using the daily average of U.S. 1-Year Treasury Bills' interest rates in 2022 and 2023. Following the prior work in \cite{yuen2001estimation}, we adopt risk aversion coefficients $\alpha_1=0.2$ and $\alpha_2=0.4$ for the retail investor and leading expert, respectively, and we also consider the case where $\alpha_1=0.4$ and $\alpha_2=0.2$. We vary the imitation coefficient $\theta$ across values of $\frac{1}{4}$, $1$, $4$, and $16$. Note that the retail investor's initial wealth can be any positive real number \cite{rogers2013optimal}, we set $X_1(0)=1$. We observe the same trend for other values of the parameters. 

\subsection{Experiment Results}
The experiment results for Case 1 and Case 2 are shown in Fig. \ref{fig:fig1} and Fig. \ref{fig:fig2}, respectively. The solid lines represent the retail investor's rational decision $\{\bar{P}_1(t)\}_{t\in\mathcal{T}}$ and the leading expert's rational decision $\{\bar{P}_2(t)\}_{t\in\mathcal{T}}$ calculated using \eqref{eq:barP1} and \eqref{eq:merton-optimal-decision}, respectively. The dashed lines represent the retail investor's optimal decisions $\{P_1^*(t)\}_{t\in\mathcal{T}}$ with different values of the imitation coefficients $\theta$ calculated using Theorem \ref{th:1}--\ref{th:2+}. The dotted lines represent the retail investor's asymptotic decisions $\{P_{1\infty}^*(t)\}_{t\in\mathcal{T}}$ calculated using Theorem \ref{th:3}. 

\subsubsection{Solution to the optimal decision}
We can obtain the real optimal decisions in Problem 2 using the numerical solver in Python. The theoretical results calculated using Theorem \ref{th:1}--\ref{th:2+} match the real results well, which validates the correctness of Theorem \ref{th:1}--\ref{th:2+}. Details are in the supplementary file.

\subsubsection{Impact of decision-changing imitation on the optimal decision}
\begin{figure}[!t]
\centering
\subfloat[$\alpha_1=0.2,\alpha_2=0.4$]{\includegraphics[width=0.35\linewidth]{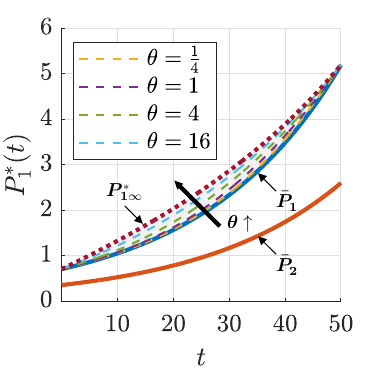} \label{fig:fig1a}}
\subfloat[$\alpha_1=0.4,\alpha_2=0.2$]{\includegraphics[width=0.35\linewidth]{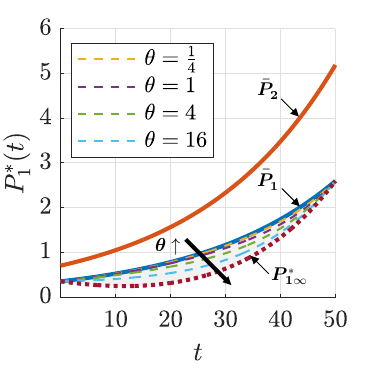} \label{fig:fig1b}}
\caption{Optimal decisions $\{P_1^*(t)\}_{t\in\mathcal{T}}$, rational decisions $\{\bar{P}_1(t)\}_{t\in\mathcal{T}}$ and $\{\bar{P}_2(t)\}_{t\in\mathcal{T}}$, and asymptotic decisions $\{P_{1\infty}^*(t)\}_{t\in\mathcal{T}}$ in Case 1.}
\label{fig:fig1}
\end{figure}


\noindent \textbf{Case 1.} From Fig. \ref{fig:fig1a}, it can be observed that when $\alpha_1<\alpha_2$, we have $P_{1\infty}^*(t)\geqslant\bar{P}_1(t)>\bar{P}_2(t)$ for $t\in\mathcal{T}$, i.e., the retail investor's asymptotic decision diverges more from the leading expert's rational decision compared to his/her rational decision, which validates Theorem \ref{co:1}. Furthermore, we find that the conclusion in Theorem \ref{co:1} is also suitable for the general cases when $\theta\in(0,+\infty)$, i.e., when $\alpha_1<\alpha_2$, we have $P_1^*(t)\geqslant\bar{P}_1(t)>\bar{P}_2(t)$ for $t\in\mathcal{T}$. From Fig. \ref{fig:fig1b}, we can draw the same conclusion when $\alpha_1>\alpha_2$. 

\begin{figure}[!t]
\centering
\subfloat[$\alpha_1=0.2,\alpha_2=0.4$]{\includegraphics[width=0.35\linewidth]{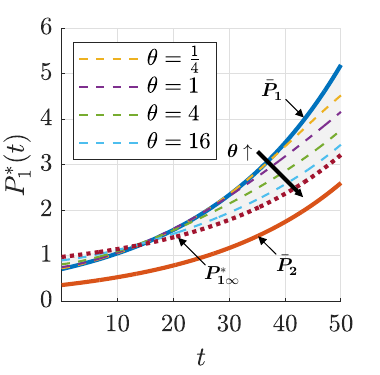} \label{fig:fig2a}}
\subfloat[$\alpha_1=0.4,\alpha_2=0.2$]{\includegraphics[width=0.35\linewidth]{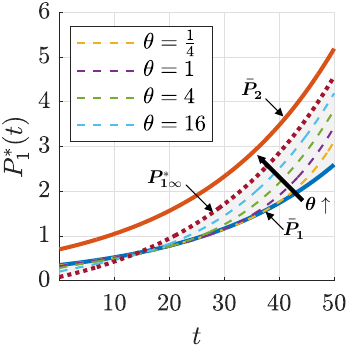} \label{fig:fig2b}}
\caption{Optimal decisions $\{P_1^*(t)\}_{t\in\mathcal{T}}$, rational decisions $\{\bar{P}_1(t)\}_{t\in\mathcal{T}}$ and $\{\bar{P}_2(t)\}_{t\in\mathcal{T}}$, and asymptotic decisions $\{P_{1\infty}^*(t)\}_{t\in\mathcal{T}}$ in Case 2.}
\label{fig:fig2}
\end{figure}


\noindent \textbf{Case 2.} From Fig. \ref{fig:fig2a}, it can be observed that there exists $\tau\in\mathcal{T}$ which satisfies $P_{1\infty}^*(\tau)=\bar{P}_1(\tau)$. When $\alpha_1<\alpha_2$, we have $P_{1\infty}^*(t)\geqslant\bar{P}_1(t)>\bar{P}_2(t)$ for $t\in[0,\tau]$, i.e., the retail investor's asymptotic decision diverges more from the leading expert's rational decision compared to his/her rational decision before $\tau$. Also, we observe $\bar{P}_1(t)\geqslant P_{1\infty}^*(t)>\bar{P}_2(t)$ for $t\in[\tau,T]$, i.e., the retail investor's asymptotic decision falls between the two investors' rational decisions after $\tau$, which validates Theorem \ref{co:2}. Furthermore, from Fig. \ref{fig:fig2a}, we find that the conclusion in Theorem \ref{co:2} is also suitable for the general cases when $\theta\in(0,+\infty)$. First, there exists $\tau(\theta)\in\mathcal{T}$ which satisfies $P_1^*(\tau(\theta))=\bar{P}_1(\tau(\theta))$. When $\alpha_1<\alpha_2$, we have $P_1^*(t)\geqslant\bar{P}_1(t)>\bar{P}_2(t)$ for $t\in[0,\tau(\theta)]$, and we have $\bar{P}_1(t)\geqslant P_1^*(t)>\bar{P}_2(t)$ for $t\in[\tau(\theta),T]$. From Fig. \ref{fig:fig2b}, we can draw the same conclusion when $\alpha_1>\alpha_2$. 

\section{Conclusion}
\label{sec:conclusion}
Decision-changing imitation is a prevalent phenomenon in financial markets. In this work, we study the optimal investment problem under the influence of decision-changing imitation involving one leading expert and one retail investor whose decisions are unilaterally influenced by the leading expert. We use the variational method to derive the general solution to the optimal investment problem and determine the retail investor's optimal decision under two special cases of the boundary conditions. Our theoretical analysis reveals that when retail investors prefer higher risk than leading experts, if they set their initial and terminal decisions as rational decisions, imitating the leading experts' decision-changing rate makes them more risk-seeking than their rational decisions. If they do not set these decisions as rational decisions, they become more risk-seeking during the earlier part of the investment period and less risk-seeking during the latter part. 


\bibliographystyle{IEEEtran}
\bibliography{mybib}

\end{document}